\documentclass[10pt,journal]{IEEEtran}
\usepackage{textcomp}
\usepackage{enumerate}
\usepackage{algorithmic}
\usepackage{amsfonts}
\usepackage{graphicx,subcaption}
\usepackage{amsmath}
\usepackage{amssymb}
\usepackage{mathrsfs}
\usepackage{cite}      % Written by Donald Arseneau
\renewcommand\citepunct{, }

\usepackage{graphicx}  % Written by David Carlisle and Sebastian Rahtz
\begin{document}
\title{A Near-optimal User Ordering Algorithm for Non-iterative Interference Alignment Transceiver Design in MIMO Interfering Broadcast Channels}
%\author{Chiao-En Chen,~\IEEEmembership{Member,~IEEE}}

\author{Chiao-En Chen,~\IEEEmembership{Senior Member,~IEEE} 
\thanks{
This work was partially supported in part by the National Science Council under the Grants NSC 102-2221-E-194-009-MY2.}
\thanks{
Chiao-En Chen is with the Department of Electrical/Communications Engineering,
National Chung Cheng University, Chiayi, Taiwan. (e-mail:
ieecec@ccu.edu.tw).}
}

\maketitle

\begin{abstract}
Interference alignment (IA) has recently emerged as a promising interference mitigation technique for interference networks. In this letter, we focus on the IA non-iterative transceiver design problem in a multiple-input-multiple-output interfering broadcast channel (MIMO-IBC), and observed that there is previously unexploited flexibility in different permutations of user ordering. By choosing a good user ordering for a pre-determined IA inter-channel-interference allocation, an improved transceiver design can be accomplished. In order to achieve a more practical performance-complexity tradeoff, a suboptimal user ordering algorithm is proposed. Simulation shows the proposed suboptimal user ordering algorithm can achieve near-optimal performance compared to the optimal ordering while exhibiting only moderate computational complexity.

\end{abstract}



% Will be ignored for other modes.
\IEEEpeerreviewmaketitle

\section{Introduction}

% \PARstart{A}{demo} file is ....
% \PARstart{A}{}demo file is ....
% \PARstart{T}{his demo} file is ....
%\PARstart{T}{his} demo file is intended to serve as a ``starter file"
%for IEEE journal papers produced under \LaTeX\ using IEEEtran.cls version
%1.6b and later.
%% You must have at least 2 lines in the paragraph with the drop letter
%% (should never be an issue)
% May all your publication endeavors be successful.
%\hfill November 18, 2002
%\PARstart{S}{patial} modulation (SM) 
%\cite{Mesleh:08_IEEE_TVT} is a recently proposed multiple-input-multiple-output (MIMO) scheme which combines the conventional amplitude/phase modulation with antenna index modulation. Compared to the conventional schemes, SM enables a higher throughput, simpler transceiver design, and better energy efficiency tradeoffs. Hence, it has drawn great research attentions lately. Many of the recent advances in SM-related techniques can be found in the survey papers \cite{Renzo:14_IEEE_proc,Yang:14_IEEE_CST} and also references therein.
%
%

\PARstart{I}{n} a multiple-input-multiple-output interfering broadcast channel (MIMO-IBC), multiple mobile stations (MSs), each equipped with multiple antennas, simultaneously receive multiple spatial streams from their serving multiple-antenna base station (BS). The channel state information of different cells are shared among the BSs to enable better interference managing strategies, which is crucial for delivering high-throughput high-reliability communications services \cite{Boudreau:09_IEEE_CM}. Interference alignment (IA) \cite{Jafar:11_book} is one of these interference managing techniques which facilitates effective interference suppression by first aligning  multiple interference signals into a subspace of dimension much smaller than the number of interferers.  While IA is originally proposed for the interference channel (IC) \cite{Cadambe:08_IEEE_TIT}, it has also recently found great use for transceiver designs in MIMO-IBC 
\cite{Shin:11_IEEE_TWC, Suh:11_IEEE_TC,Tang:13_IEEE_TC,Liu:12_WCNC,Lee:14_IEEE_TWC}, and other types of MIMO channels.

IA transceiver design in MIMO-IBC is much more challenging than that in the MIMO-IC because not only the inter-cell interference (ICI) and inter-stream interference (ISI) but also the inter-user-interference (IUI) have to be taken into account. Some early studies of this problem have been restricted to the two-cell scenario \cite{Shin:11_IEEE_TWC, Suh:11_IEEE_TC}. In \cite{Shin:11_IEEE_TWC}, a closed-form solution has been derived for a two-cell IBC in which each cell can only supports two users. The downlink interference alignment scheme \cite{Suh:11_IEEE_TC} features a within-a-cell feedback mechanism and allows more than two users per-cell, but still cannot be applied to cases with more than two cells. Recently, IA transceiver designs for the general multi-cell MIMO-IBC have also been reported. In \cite{Tang:13_IEEE_TC}, the authors proposed a grouping method which can extend the IA design in \cite{Shin:11_IEEE_TWC} to the multi-cell scenario. In \cite{Liu:12_WCNC}, the authors proposed two different transceiver designs (case-I and case-II) for the multi-cell MIMO-IBC, although the case-II design does not necessarily converge to an IA solution \cite{Liu:12_WCNC}. More recently, a non-iterative IA transceiver design has also been proposed for a multi-cell MIMO-IBC \cite{Lee:14_IEEE_TWC}. Compared to the design in \cite{Tang:13_IEEE_TC}, \cite{Lee:14_IEEE_TWC} requires a smaller number of antennas to achieve the same total degrees-of-freedom (DOF) and hence is more advantageous from the hardware perspective. In contrast to the IA case-I design of \cite{Liu:12_WCNC} which requires the MS having more antennas than the BS, the design in \cite{Lee:14_IEEE_TWC} avoids this restriction and hence is more realistic in practical cellular networks.

In spite of the efficiency and practicality, we first note that there is previously unexploited flexibility in different permutations of user ordering in the transceiver design proposed in \cite{Lee:14_IEEE_TWC}. By choosing a good user ordering for a pre-determined IA ICI allocation, improved transceiver design can be accomplished. It is worthwhile noting that exploiting user ordering to improve the system's performance is not a new concept for MIMO systems when successive interference cancellation (SIC) detection or pre-cancellation (precoding) \cite{Wolniansky:98_URSI,Liu:07:CJECE} is used. However, to the best of the author's knowledge, the idea of exploiting user ordering for IA linear transceiver design in a MIMO-IBC setting is completely new and has not been reported in the literature.

Clearly, to achieve the best performance, the optimal user ordering can be obtained by the brute-force exhaustive search which is computationally prohibitive when the search space is large. Aiming at achieving a more balanced performance-complexity tradeoff, we propose a suboptimal user ordering algorithm based on the technique of alternating optimization. Simulation shows the proposed suboptimal user ordering substantially improves the performance of the original IA transceiver design \cite{Lee:14_IEEE_TWC}, and exhibits near-optimal performance with significantly lower computational complexity.

\textit{Notations:} Throughout this letter, matrices and vectors are
set in boldface, with uppercase letters for matrices and lower case
letters for vectors. We use the symbol $\boldsymbol{\pi}=\{\pi(1),\ldots, \pi(K)\}$ to represent an ordered set of $K$ elements with $\pi(i)$ being the $i$th element. The superscript $^\mathrm{H}$ denotes the conjugate transpose, $\mathrm{E}\{\cdot\}$ is the expectation operator. $\|\cdot\|$ denotes the Euclidean norm of a vector, and $\mathbf{I}_N$ denotes the $N\times N$ identity matrix.

%$\mathrm{tr}\{\cdot\}$ and $\mathrm{det}\{\cdot\}$ denote
%the trace and determinant of a matrix, respectively. 
%
%
%$\mathrm{E}\{\cdot\}$ denotes the expectation operator and $\Re\{\mathbf{X}\}$ denote
%the real part of $\mathbf{X}$. The superscripts $^T$, $^H$, and $^{\dagger}$ denote the
%transpose, conjugate transpose, and the right-inverse of a matrix respectively. For a full row rank matrix $\mathbf{A}$, $\mathbf{A}^{\dagger}=\mathbf{A}^H(\mathbf{A}\mathbf{A}^H)^{-1}$. $\mathbf{I}_N$ denotes the $N\times N$ identity matrix.

\section{System Model}
\label{eq:system}
We consider a $C$-cell MIMO-IBC in which the $i$th cell consists of a BS and $K_i$ MSs. The BS in the $i$th cell is assumed to be equipped with $M_i$ antennas, while each MS in the $i$th cell is assumed to receive $d$ spatial streams using its $N_i$ antennas ($N_i\leq M_i$), for all $i=1,\ldots, C$. For convenience of later discussion, we use the shorthand notation $(C, [K_1,\ldots, K_C], d, \{M_1,\ldots, M_C/N_1,\ldots, N_C\})$ as in \cite{Lee:14_IEEE_TWC} to represent the system configuration mentioned above.

Following the typical assumption that the channel is frequency flat and i.i.d. (independent and identically distributed) Rayleigh faded, the channel from the $i$th BS to the $[k,j]$th MS (the $k$th MS in the $j$th cell) can be modelled by the channel matrix $\mathbf{H}_i^{[k,j]}\in \mathbb{C}^{N_j\times M_i}$. Let $\mathbf{x}^{[k,i]}\in \mathbb{C}^{d\times 1}$ be the normalized symbol vector targeted for the $[k,i]$th MS, where $\mathrm{E}\{\mathbf{x}^{[k,i]}\mathbf{x}^{[k,i]}{}^{\mathrm{H}}\}=\mathbf{I}_d$, and $\mathbf{V}^{[k,i]}=[\mathbf{v}^{[k,i]}_1,\ldots, \mathbf{v}^{[k,i]}_d]\in \mathbb{C}^{M_i\times d}$ be the associated precoding matrix, then the data vector at the $[k,i]$th MS can be described as
\begin{align}
\mathbf{y}^{[k,i]}=&\mathbf{H}_i^{[k,i]}\mathbf{V}^{[k,i]}\mathbf{x}^{[k,i]}+\sum_{\substack{\ell=1\\ \ell\neq k}}^{K_i}\mathbf{H}_i^{[k,i]}\mathbf{V}^{[\ell,i]}\mathbf{x}^{[\ell,i]}\nonumber\\
+&
\sum_{\substack{j=1\\j\neq i}}^C\sum_{\ell=1}^{K_j}\mathbf{H}_j^{[k,i]}\mathbf{V}^{[\ell,j]}\mathbf{x}^{[\ell,j]}+\mathbf{n}^{[k,i]}.
\end{align}
Here $\mathbf{n}^{[k,i]}\in \mathbb{C}^{N_i\times 1}$ denotes the noise vector at the $[k,i]$th MS, which is modelled as a circularly symmetric complex Gaussian vector with zero mean and covariance matrix $\sigma^2\mathbf{I}_{N_i}$. We assume the transmitted signal at the $i$th BS satisfies the total power constraint $\sum_{k=1}^{K_i}\mathrm{E}\{\|\mathbf{V}^{[k,i]}\mathbf{x}^{[k,i]}\|^2\}\leq P$. After receiving the data vector, the $[k,i]$th MS then applies a linear combining matrix $\mathbf{U}^{[k,i]}{}^{\mathrm{H}}=[\mathbf{u}^{[k,i]}_1,\ldots, \mathbf{u}^{[k,i]}_d]^{\mathrm{H}} \in \mathbb{C}^{d \times N_i}$ to $\mathbf{y}^{[k,i]}$ for signal enhancement and interference/noise suppression. The achievable sum rate of the $[k,i]$th MS is given by
\begin{align}
R^{[k,i]}=\sum_{m=1}^{d}\log_2\left(1+\frac{\left|\mathbf{u}_m^{[k,i]}{}^{\mathrm{H}}\mathbf{H}_{i}^{[k,i]}\mathbf{v}_m^{[k,i]}\right|^2}{\sigma^2\|\mathbf{u}_m^{[k,i]}\|^2+I_m^{[k,i]}}\right),
\label{eq:R_ki}
\end{align}
where
\begin{align}
I_{m}^{[k,i]}=&\sum_{\substack{n=1\\n\neq m}}^d \left|\mathbf{u}_m^{[k,i]}{}^\mathrm{H}\mathbf{H}_i^{[k,i]}\mathbf{v}_n^{[k,i]}\right|^2+\sum_{\substack{\ell=1 \\ \ell \neq k}}^{K_i}\sum_{n=1}^{d} \left|\mathbf{u}_m^{[k,i]}{}^\mathrm{H}\mathbf{H}_i^{[k,i]}\mathbf{v}_n^{[\ell,i]}\right|^2\nonumber\\
+& \sum_{\substack{j=1\\j\neq i}}^C\sum_{\ell=1}^{K_j}\sum_{n=1}^{d} 
\left|\mathbf{u}_m^{[k,i]}{}^\mathrm{H}\mathbf{H}_j^{[k,i]}\mathbf{v}_n^{[\ell,j]}\right|^2.
\end{align}

\section{Overview on non-iterative IA transceiver design for MIMO-IBC}
\label{sec:overview}
The design procedure for the MIMO-IBC non-iterative IA transceiver proposed in  \cite{Lee:14_IEEE_TWC} can be summarized as follows:
\begin{enumerate}[Step 1.]
\item Given $C$ and $K_j$, the total number of effective ICI channels for the users in the $j$th cell, denoted by $S_j$, is determined from the the lower bound $(C-2)K_j+1\leq S_j$, for all $j=1,\ldots, C$.

\item Given $C$, $\{K_i\}$, and $S_j$, determine the number of the effective ICI channels from the $i$th BS ($i\neq j$) to the users in the $j$th cell, $t_{j,i}$. Let the $m$th basis vector of the $s$th effective ICI channel from the $i$th BS to the users in the $j$th cell be denoted as $\mathbf{q}_{j,i,m}^{(s)}$, for all $s=1,\ldots, t_{j,i}$. The number of effective ICI channels to be aligned to $\mathbf{q}_{j,i,m}^{(s)}$, denoted by $n_{j,i}^{(s)}$, is then determined by a carefully designed ICI channel allocation algorithm \cite{Lee:14_IEEE_TWC}. 
\item The BS computes $\{\mathbf{q}_{j,i,m}^{(s)}\}$ and receive beamforming vectors $\{\mathbf{u}_{m}^{[k,j]}\}$ such that $n_{j,i}^{(s)}$ effective ICI channels are perfectly aligned to $\mathbf{q}_{j,i,m}^{(s)}$. This can be achieved by solving a matrix equation which involves the information of  $\{\mathbf{H}_{i}^{[k,j]}\}$, $\{n_{j,i}^{(s)}\}$, and $\{t_{j,i}\}$.
\item
The BS then computes the transmit beamforming vectors $\mathbf{v}_{m}^{[k,i]}$ such that $\mathbf{v}_{m}^{[k,i]}\subset \text{null}([\text{IUI,\;ICI,\;ISI}]^{\mathrm{H}})$, where
\begin{align}
\text{IUI}&=\left[\mathbf{W}_{i}^{[1,i]},\ldots,\mathbf{W}_{i}^{[k-1,i]},\mathbf{W}_{i}^{[k+1,i]},\ldots, \mathbf{W}_{i}^{[K_i,i]}\right],\nonumber\\
\text{ICI}&=\left[\mathbf{Q}_{1,i}^{(1)},\ldots,\mathbf{Q}_{i-1,i}^{(t_{i-1,i})},\mathbf{Q}_{i+1,i}^{(1)},\ldots, \mathbf{Q}_{C,i}^{(t_{C,i})}\right],\nonumber\\
\text{ISI}&=\left[\mathbf{w}_{i,1}^{[k,i]},\ldots,\mathbf{w}_{i,m-1}^{[k,i]},\mathbf{w}_{i,m+1}^{[k,i]},\ldots,\mathbf{w}_{i,d}^{[k,i]}\right].\nonumber
\end{align}
Here we denote $\mathbf{w}^{[k,j]}_{i,m}=\mathbf{H}_{i}^{[k,j]}{}^{\mathrm{H}}\mathbf{u}_m^{[k,j]}$, $\mathbf{W}_{i}^{[k,j]}=[\mathbf{w}^{[k,j]}_{i,1},\ldots,\mathbf{w}^{[k,j]}_{i,d}]$, and $\mathbf{Q}^{(s)}_{j,i}=[\mathbf{q}_{j,i,1}^{(s)},\ldots,\mathbf{q}_{j,i,d}^{(s)}]$.
\end{enumerate}

\section{User ordering in the non-iterative MIMO-IBC IA transceiver design}
\label{eq:user_ordering}

\subsection{Key observation}
\begin{figure}
\centering
\includegraphics[width=0.45\textwidth]{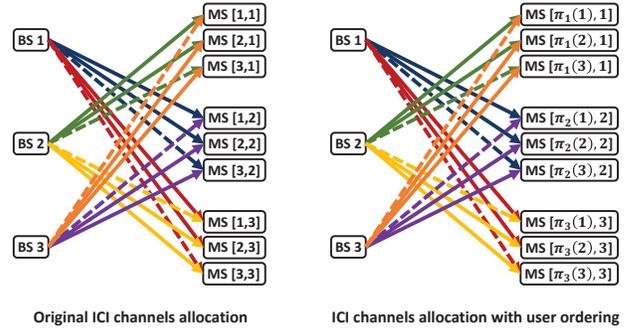}
\caption{Original and proposed ICI channel allocation for a MIMO-IBC with configuration $(3,[3,3,3],1,7,7,7/5,5,5])$. } 
\label{fig:pairing}
\end{figure}

The key observation in this paper is to first note that there is previously unexploited flexibility in different permutations of user ordering in the transceiver design of 
\cite{Lee:14_IEEE_TWC}. Fig. \ref{fig:pairing} shows the ICI channels allocation for a MIMO-IBC with configuration $(3,[3,3,3],1,
7,7,7/5,5,5])$, where the ICI channels represented by the solid lines of the same color will be aligned into the same effective ICI channel. These ICI channels are allocated and aligned according to the interference alignment conditions in order to guarantee the existence of the receive beamformers, which 
requires the MSs associated with a set of solid lines with the same color are not associated with another set of solid lines with different colors. We observe that since each MS in the same cell receives the same number of spatial streams and is equipped with the same number of antennas, introducing permutations in user ordering within each cell does not violate the interference alignment conditions achieved by the original ICI channel allocation. This idea is made clear as shown in Fig. \ref{fig:pairing}, where we use $\boldsymbol{\pi}_j=\{\pi_j(1),\pi_j(2),\ldots, \pi_j(K_j)\}$ to denote some ordered set arranged from the index set $\{1,\ldots, K_j\}$,
representing some possible ordering for the users in the $j$th cell.

Following this line, it is clear that for a general $(C, [K_1,\ldots, K_C], d, \{M_1,\ldots, M_C/N_1,\ldots, N_C\})$ MIMO-IBC, $K_j!$ different orderings $\{\pi_j(1),\pi_j(2),\ldots,\pi_j(K_j)\}$ can be arranged from the index set $\{1,\ldots, K_j\}$ for the users in the $j$th cell, for all $j=1,\ldots, C$. Since each ordering is associated with a distinct IA transceiver design, one can therefore exploit this flexibility by choosing the best transceiver design out of the $K_1!\times K_2!\times \ldots \times K_C!$ candidates. This corresponds to the optimal ordering with the exhaustive search implementation, which can be computationally prohibitive for problems with a large search space. In order to achieve a more balanced performance-complexity tradeoff, we propose a computationally efficient sub-optimal user-ordering algorithm described in the following subsection.

\subsection{Proposed sub-optimal user ordering algorithm}
\textit{\textbf{Initialization:}} The proposed algorithm is initialized by the original non-iterative IA transceiver design with natural ordering as described in \cite{Lee:14_IEEE_TWC}. Consequently, the ordered set associated to the user indices in the $i$th cell is initialized to $\boldsymbol{\pi}_i=\{1,\ldots, K_i\}$, for all $i=1,\ldots, C$. After all the transmit and receive beamforming vectors are obtained as in \cite{Lee:14_IEEE_TWC}, the performance measure $J^{(0)}$ depending on the optimization criterion is computed, and the iteration index $\theta$ is set to $1$.

\textit{\textbf{Iterations:}}
Suppose the algorithm proceeds to the $\theta$th iteration, where $\theta=\lambda C+j$ for some integers $\lambda\geq 0$ and $1\leq j \leq C$. The algorithm then generates $N_j!$ candidates of $\boldsymbol{\pi}_j$, each corresponds to a unique permutation of the set $\{1,2,\ldots, N_j\}$. Each candidate $\boldsymbol{\pi}_j$ along with $\{\boldsymbol{\pi}_1,\ldots, \boldsymbol{\pi}_{j-1},\boldsymbol{\pi}_{j+1},\ldots, \boldsymbol{\pi}_C\}$ obtained from the last iteration all together determines a unique IA transceiver design, which can be easily obtained by performing the design procedure as described in Step 3 and Step 4 of Section \ref{sec:overview} with $\{\mathbf{H}_{i}^{[k,j]}\}$ replaced by $\{\mathbf{H}_{i}^{[\pi_j(k),j]}\}$, for all $i,j=1,\ldots, C$ and $k=1,\ldots, K_j$. Among the $K_j!$ candidates of $\boldsymbol{\pi}_j$, the one that provides the best performance measure will be chosen and stored as $\boldsymbol{\pi}_j$ for the subsequent iterations. The achieved performance measure will also be recorded as $J^{(\theta)}$. The algorithm increases the 
iteration index $\theta$ by $1$, and then enters the next iteration.

The proposed algorithm keeps iterating until $\{\boldsymbol{\pi}_1,\ldots,\boldsymbol{\pi}_C\}$ has remained fixed over $C$ consecutive iterations, or after $\theta$ has reached a pre-determined number of iterations. 

\textit{Remark 1:} It is worthwhile noting that the proposed user-ordering algorithm is very general and can be used in conjunction with all sorts of optimization criteria. In this letter, we consider both the max sum-rate criterion which maximizes the overall sum-rate $J^{(0)}=\sum_{i=1}^C\sum_{k=1}^{K_i}R^{[k,i]}$, and also the 
the max-min sum-rate criterion which aims at maximizing the minimum sum-rate $J^{(0)}=\min_{i=1,\ldots,C,\;k=1:\ldots,K_i}R^{[k,i]}$ among all MSs in the system.

\textit{Remark 2:} It is not difficult to see that $\{J^{(\theta)}\}$ is a monotonic sequence. Since the total number of possible IA transceiver designs is finite ($K_1!\times K_2!\times \ldots \times K_C!$), $\{J^{(\theta)}\}$ cannot diverge and hence the algorithm is guaranteed to terminate in finite number of iterations.

\textit{Complexity Analysis:} The required complexity in user ordering algorithm can be evaluated by the number of candidate IA transceivers being tested. For the optimal exhaustive search algorithm, a fixed number $K_1!\times K_2!\times \ldots \times K_C!$  candidates have to be tested. On the other hand, the number of IA candidates to be tested in the proposed suboptimal user ordering algorithm is a random number which depends on the channel realization and also the signal-to-noise ratio (SNR). Suppose the algorithm terminates in $\theta_{\mathrm{term}}$ iterations, where $\theta_{\mathrm{term}}=\lambda_{\mathrm{term}}C+j_{\mathrm{term}}$, then the total number of candidates being tested can be easily shown to be 
$1+\lambda_{\mathrm{term}}\sum_{i=1}^{C}K_i!+\sum_{i=1}^{j_\mathrm{term}}K_i!$.

\section{Simulation Results}
\label{sec:simulation}
In this section, we present the simulation results of the proposed suboptimal user ordering algorithm applied to the non-iterative IA transceiver design \cite{Lee:14_IEEE_TWC} under a variety of system configurations. Throughout the simulation, all the channels are assumed to be frequency flat Rayleigh fading, with each entry drawn i.i.d. from a complex Gaussian distribution with unit variance. To highlight the performance advantages, the system's overall sum-rate is plotted for comparison when the max sum-rate criterion is being used, while the sum-rate of the MS with lowest sum-rate is plotted for comparison when the maxmin sum-rate criterion is being used. The SNR is defined as $\mathrm{SNR}=P/\sigma^2$. Each simulation point is obtained by averaging over $1000$ independent channel realizations.

\begin{figure}[bht]
\centering
\includegraphics[width=0.45\textwidth]{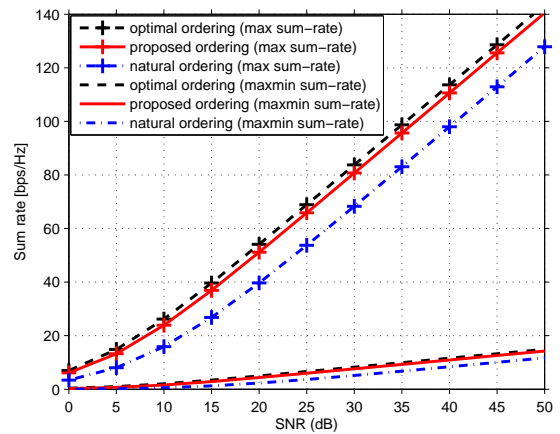}
\caption{Comparison of sum rate performance achieved by a number of ordering algorithms in a MIMO-IBC with configuration $(3,[3,3,3],1,7,7,7/5,5,5])$.} 
\label{fig:max_sumrate_fig_5}
\end{figure}

In Fig. \ref{fig:max_sumrate_fig_5}, the sum rate performance achieved by various ordering algorithms under system configuration $(3,[3,3,3],1,7,7,7/5,5,5])$ have been demonstrated. Under the max sum-rate criterion, simulation shows the proposed suboptimal ordering algorithm can provide roughly $4.3$ dB performance gain compared to the original IA transceiver design with natural ordering, and performs very close to the optimal ordering algorithm with a performance loss no more than $0.9$ dB for sufficiently high SNR. Under the max-min sum-rate criterion, the proposed suboptimal ordering algorithm performs very close to the optimal ordering and exhibits roughly $7.5$ dB gain over the original design.

\begin{figure}
\centering
\includegraphics[width=0.45\textwidth]{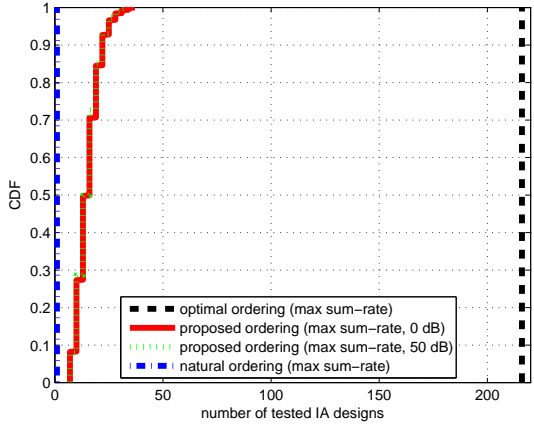}
\caption{Empirical CDFs of the number of tested IA designs required
in various user ordering algorithms under configuration $(3,[3,3,3],1,7,7,7/5,5,5])$.} 
\label{fig:complexity_max_sumrate_fig_5}
\end{figure}

\begin{figure}
\centering
\includegraphics[width=0.45\textwidth]{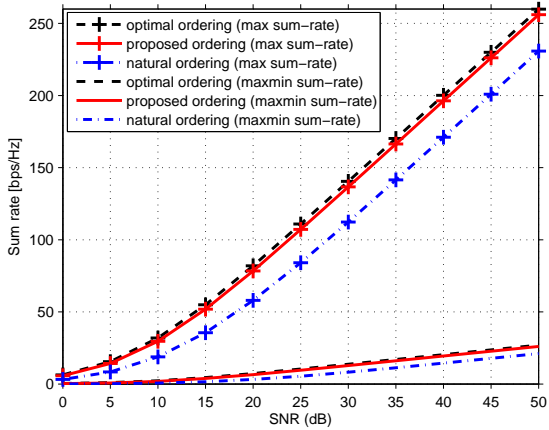}
\caption{Comparison of sum rate performance achieved by a number of ordering algorithms in a MIMO-IBC with configuration $(3,[3,2,4],2,14,12,16/10,8,10])$.} 
\label{fig:max_sumrate_fig_4}
\end{figure}

In Fig. \ref{fig:complexity_max_sumrate_fig_5}, we compare the complexity of various ordering algorithms under the max sum-rate criterion by showing the empirical cumulative distribution function (CDF) of the number of tested IA designs over $1000$ channel realizations under the same configuration as in Fig. \ref{fig:max_sumrate_fig_5}. In this setting, the optimal ordering algorithm exhaustively searches over $(3!)^3=216$ candidate IA designs and chooses the one that maximizes the system's sum-rate, while the original IA with natural ordering only implements single IA design. Simulation shows that the proposed suboptimal ordering algorithm only examines less than $22$ IA designs for $90\%$ of the channel realizations, and does not examine more than $40$ IA designs over the simulated $1000$ channel realizations. It is also observed that the complexity of the proposed suboptimal ordering algorithm appears to be not very sensitive to the SNR, as the simulation shows the difference between the complexity results simulated under $0$ dB and $50$ dB is very small.

In Fig. \ref{fig:max_sumrate_fig_4}, we compare the sum rate performance achieved by various ordering algorithms under a different system configuration $(3,[3,2,4],2,14,12,16/10,8,10])$, in which every MS now receives $2$ spatial streams from its serving BS. Simulation shows the sum rate performance achieved by the proposed suboptimal ordering algorithm under the max sum-rate criterion is only $0.4$ dB lower than the optimal ordering algorithm at high SNR, and can provide roughly $4.1$ dB gain over the original IA with natural ordering. Under the max-min sum-rate criterion, simulation shows the proposed suboptimal ordering performs very close to the optimal ordering and provides roughly $7.6$ dB gain over the original design.

\section{Conclusion}
\label{eq:conclusion}
In this letter, we first showed that there is previously unexploited flexibility in user ordering in the recently proposed non-iterative IA transceiver design for MIMO-IBC. In additional to the optimal ordering using exhaustive search, a suboptimal user ordering algorithm is also proposed to achieve a more practical performance-complexity tradeoff. Simulation results show that the proposed algorithm can provide substantial performance gain compared to the original non-iterative IA transceiver design and achieve near-optimal performance with significantly lower computational complexity.

\end{document}